\documentclass[3p]{elsarticle}

\usepackage{graphicx}
\usepackage{physics}
\usepackage{amsfonts}
\usepackage{subcaption}
\usepackage[colorlinks=true]{hyperref}

\newcommand{\sS}{\textsc{s}}
\newcommand{\sB}{\textsc{b}}
\newcommand{\sG}{\textsc{g}}
\newcommand{\sI}{\textsc{i}}
\newcommand{\sQ}{\textsc{q}}
\newcommand{\sW}{\textsc{w}}
\newcommand{\sSB}{\textsc{sb}}
\newcommand{\sQV}{\textsc{qv}}
\newcommand{\sSV}{\textsc{sv}}

\biboptions{sort&compress}

\begin{document}

\title{Entropy production  of a small quantum system under strong coupling with an environment: A computational experiment}

\author[uab]{Ketan~Goyal}
\author[uab,nanchan]{Xian~He}
\author[uab]{Ryoichi~Kawai\corref{cor1}}

\address[uab]{Department of Physics, University of Alabama at Birmingham, Birmingham, AL 35294, USA}
\address[nanchan]{Department of Physics, Nanchang University, Nanchang 330031, China}
\cortext[cor1]{Corresponding author}

\begin{abstract}
Many theoretical expressions of dissipation along non-equilibrium processes have been proposed.  However, they have not been fully verified by experiments. Especially for systems strongly interacting with environments the connection between theoretical quantities and standard thermodynamic observables are not clear. We have developed a computer simulation based on a spin-boson model, which is in principle exact and suitable for testing the proposed theories.  We have noted that the dissipation obtained by measuring conventional thermodynamic quantities deviates from the second law of thermodynamics presumably due to the strong coupling.  We show that additive correction to entropy makes it more consistent with the second law.  This observation appears to be consistent with the theory based on the potential of mean force.
\end{abstract}
\begin{keyword}
irreversibility \sep entropy production \sep strong coupling \sep system-environment correlation \sep quantum thermodynamics \sep non-Markovian dynamics \sep hierarchical equation of motion
\end{keyword}

\maketitle 
\centerline{
This work is dedicated for the memory of Christian Van den Broeck.}

\section{Introduction}

The second law of thermodynamics is considered to be one of the most fundamental laws of physics.\cite{Eddington1927} Yet the precise definition of irreversible entropy production still remains elusive.  It is now widely accepted that von Neumann entropy (vN-entropy), $S(\rho) = - \tr \rho \ln \rho$ where $\rho$ is the density operator of the system, represents thermodynamic entropy even in non-equilibrium situations.  However, it is invariant under unitary evolution owing to the Liouville theorem and the microscopic time reversibility, and thus the entropy of isolated systems never changes.  On the other hand, the second law and entropy production can be determined by comparing forward and time-reversed processes\cite{Kawai2007,Parrondo2009}, suggesting that the irreversibility of thermodynamics systems  does not conflict with the microscopic time reversibility. Esposito-Lindenberg-Van den Broeck(ELV) has recently developed a general ``non-equilibrium'' expression of entropy production based on the vN-entropy\cite{Esposito2010},  which is consistent with fluctuation theorems.\cite{Funo2019} 
While the proposed expression is developed under a rather general condition, its validity has not been tested, in particular for systems strongly interacting with environments.

A standard theory of thermodynamics assumes that interaction energy between system and environment is negligibly small compared to the system energy. Under such a condition, boarder lines between a system and environments are obvious and the thermodynamic laws can be expressed only with the state of system. However, the assumption of weak coupling becomes untenable when the system is reduced to a molecular size where the system energy and the coupling energy are of the same order. Then, the separation becomes moot since the interaction energy depends on the state of system and environment simultaneously. Because of this inseparability, standard thermodynamic quantities such as heat and entropy are no longer clearly defined. It is natural to ask if the thermodynamic laws can be expressed only with the state of system or the state of environments must be taken into account.  Many works have been reported\cite{Gelin2009,Campisi2010,Esposito2010,Kim2010,Hilt2011,Gallego2014,Esposito2015,Wang2015,Gelbwaser-Klimovsky2015,Seifert2016,Katz2016,Uzdin2016,Carrega2016,Jarzynski2017,Miller2017,Newman2017,Aurell2017,Dou2018,Bruch2018,Perarnau-Llobet2018,Xu2018,Funo2018,Strasberg2019,Funo2019}  but the issues have not been resolved yet. 

In addition to the issue of interaction energy, the correlation between system and environment, especially quantum entanglement, plays an important role in thermodynamics.  For example, the coupling introduces irreversible change in the information entropy through decoherence induced by quantum entanglement between system and environments\cite{Zurek2003,Schlosshauser2007}.  Moreover, under the strong coupling limit, the system density no longer takes a canonical form \cite{Lee2012,Iles-Smith2014,Goyal2017} and even not diagonal in the energy basis.\cite{Campisi2009,Hilt2011,Goyal2017} It has been suggested that continuous measurement of the system by the environments projects the Gibbs state onto the so-called pointer basis.\cite{Goyal2017} Continuous measurement also suppresses heat conduction due to the quantum Zeno effect under the strong coupling limit.\cite{Kato2015,Kato2016,Goyal2017,Maier2019}. Since the quantum coherency and entanglement can be used as resources in thermodynamics\cite{Sagawa2008,Sagawa2009,Toyabe2010,Funo2018}, the decoherence should be considered as a part of thermodynamic dissipation.

A promising method that takes into account strong coupling was developed long ago by  Kirkwood\cite{Kirkwood1935}, in which an effective Hamiltonian or so-called potential of mean force, replaces the Hamiltonian of system. In this approach, the effects of strong coupling are incorporated into effective thermodynamic quantities such that they satisfy the standard thermodynamics laws. The same idea has recently been extended to stochastic thermodynamics.\cite{Gelin2009,Campisi2009,Campisi2010,Hilt2011,Esposito2015,Seifert2016,Uzdin2016,Jarzynski2017,Miller2017,Strasberg2019}.  In particular, the first and second laws of thermodynamics are redefined with the effective quantities\cite{Seifert2016,Jarzynski2017,Strasberg2019}.   In this theory, the laws of thermodynamics with the effective thermodynamics quantities are still determined only by the state of system.  However, the physical meaning of effective thermodynamic quantities  and their relation to the observable physical quantities are not clear.

The present investigation tries to find where the standard theory of thermodynamics fails due to strong coupling through exact numerical experiments. We observed that the second law is violated if the conventional definition of thermodynamic variables are used.  On the other hand, we found that if an appropriate additive correction applied to entropy and other quantities, the entropy production appears to be consistent with the second law even under strong coupling. In the following, first we introduce an experimental strategy to measure necessary information from computational experiments. Then, a simple model and numerical methods are introduced.   Experiments are carried out for various different cases to ensure that the observed results are not specific to a single case. At the end, we will compare the results with a theory based on the potential of mean force.

\section{Construction of experiments}

We first construct experiments that provide accurate information about dissipation under strong coupling conditions. Consider an isolated composite system consisting of a small systems S and a large environment B.  When S is brought into contact with B, dissipation takes place. We would like to determine the irreversible entropy production by measurement only with the knowledge of conventional thermodynamics.

Hamiltonian of the whole system is given by
\begin{equation}
   H_\sSB = H_\sS + H_\sB + \lambda(t) H_\sI
\end{equation}
where $H_\sS$ and $H_\sB$ are Hamiltonians for a system and an environment, respectively.  The coupling Hamiltonian $H_\sI$ is switched on and off by a protocol $1\ge\lambda(t)\ge0$.
We measure the following energies in numerical experiments.
System energy is simply the mean energy of system:
   \begin{equation}\label{eq:def_U}
      U(t) = \tr_\sS \{\rho_\sS(t) H_\sS\}.
   \end{equation}
Heat is defined as the energy released by the environment, 
   \begin{equation}\label{eq:def_Q}
      Q(t) = \tr_\sB \{\rho_\sB(t_0) H_\sB\} - \tr_\sB \{\rho_\sB(t) H_\sB\},
   \end{equation}
and work is the change in total energy
\begin{equation}\label{eq:def_W}
      W(t) = \tr_\sSB \{\rho_\sSB(t) H_\sSB\} - \tr_\sSB \{\rho_\sSB(t_0) H_\sSB\}
\end{equation}
where $\rho_\sSB$ is the density operator of the whole system. The reduced densities $\rho_\sS=\tr_\sB \rho_\sSB$ and $\rho_\sB = \tr_\sS \rho_\sSB$ represent the state of system and environment, respectively.
In addition, we measure the coupling energy: 
\begin{equation}\label{eq:def_VI}
   V_\sI(t) = \lambda(t) \tr_\sSB\{\rho_\sSB(t) H_\sI\}.
\end{equation}

We assume that the initial state is a product state $\rho_\sSB(t_0) = \rho_\sS(t_0) \otimes \rho_\sB^\sG$ where the environment is in a local Gibbs state $\rho^\sG_\sB = e^{-\beta H_\sB}/\tr_\sB e^{-\beta H_\sB}$.  The initial state $\rho_\sS(t_0)$ is arbitrary. We will make it sure that $V_\sI(t_0)=0$,  which prevents any discontinuity in the evolution of work and avoids unnecessary dissipation. Our goal is to determine entropy production along non-equilibrium transformations between the two states, $\lambda=0$ and $\lambda=1$. 
Based on the standard theory of thermodynamics, internal energy $\mathcal{U}$, entropy $\mathcal{S}$ and free energy $\mathcal{F}$ are state functions at thermal equilibrium.  There are two ways to determine the entropy production, one from work and the other from heat as 
\begin{eqnarray}
\Sigma_\sQ &=& \Delta \mathcal{S} - \beta Q \label{eq:def_dissipation_Q} \\
\Sigma_\sW &=& W - \Delta \mathcal{F} \label{eq:def_dissipation_W}
\end{eqnarray}
where free energy difference are defined by $\Delta \mathcal{F} = \mathcal{F}(\lambda=1)-  \mathcal{F}(\lambda=0)$ and entropy difference by $\Delta \mathcal{S} = \mathcal{S}(\lambda=1) - \mathcal{S}(\lambda=0)$. Strictly speaking, the two expressions of entropy production are  valid only for a transition between two equilibriums and should take the same value.  

First, we try to find information on equilibrium states, namely $\Delta \mathcal{F}$ and $\Delta \mathcal{S}$.  Since the initial state is not at an equilibrium, we must relax it by connecting the system to the environment. For this purpose, we use a ``quasi static'' protocol (protocol 1):
\begin{equation}
   \lambda_1(t) = 
   \begin{cases}
      \text{slowly turn on coupling} & t_1 > t > t_0 \\
      \text{keep coupling constant} & t_2 > t > t_1 \\
      \text{slowly turn off coupling} & t_3 > t>  t_2 \\
      \text{keep coupling off} & t_4 > t > t_3
   \end{cases}
\end{equation}
Each step takes a sufficiently long time  that no unwanted dissipation takes place. 
During the first period, the initial non-equilibrium state relaxes to an equilibrium and the initial dissipation  $\Sigma^0$ completes before $t_1$. At the same time the system energy also relaxes by $\Delta U^0$. No further dissipation takes place after $t_1$.   Then, the measurements of heat and work at $t_2$ and $t_4$ provide information on the equilibrium states.  In real experiments, the two measurements must be done in separate experiments since the measurement at $t_2$ changes the quantum states afterward.

Since $H_\sSB(t_0)=H_\sSB(t_4)$, we have $\Delta \mathcal{S}(t_4) =\mathcal{S}(t_4)-\mathcal{S}(t_0)=0$. Hence, net heat observed at $t_4$ is entirely dissipative and entropy production due to the relaxation of initial non-equilibrium state is given by
\begin{equation}\label{eq:sigma0_Q4}
\Sigma^0_\sQ(t_4) = -\beta Q^\text{qs}(t_4)
\end{equation} 
Similarly we have $\Delta \mathcal{F}(t_4) =\mathcal{F}(t_4)-\mathcal{F}(t_0)=0$. However, we must take into account the free energy difference from the initial non-equilibrium state to the final equilibrium,  which we assume $\Delta F^0 \equiv \Delta U^\text{qs}(t_4) = U^\text{qs}(t_4)-U^\text{qs}(t_0)$.
Then, the dissipation obtained from work is 
\begin{equation}\label{eq:sigma0_W4}
\Sigma^0_\sW= \beta \left[ W^\text{qs}(t_4)  - \Delta U^\text{qs}(t_4)\right]
\end{equation}
The energy conservation law $W^\text{qs}(t_4)+Q^\text{qs}(t_4) = \Delta U^\text{qs}(t_4)$ guarantees that the two expressions of entropy production (\ref{eq:sigma_Q4}) and (\ref{eq:sigma_W4}) are equivalent. 

Next we determine $\Delta \mathcal{F}$ and $\Delta \mathcal{S}$ by measuring $W$ and $Q$ at $t_2$ by which the whole system reaches the equilibrium with $\lambda=1$.  The free energy and entropy differences between the fully detached and connected states are determined as
\begin{equation}\label{eq:DF}
   \Delta \mathcal{F} = W^\text{qs}(t_2) - W^\text{qs}(t_4) + \Delta U^\text{qs}(t_4).  
\end{equation}
\begin{equation}\label{eq:DS}
   \Delta \mathcal{S} = \beta \left[Q^\text{qs}(t_2) - Q^\text{qs}(t_4)\right].
\end{equation} 
From the theoretical view, these expressions look obvious.  However, since we do not know the equilibrium state at the strong coupling this measurement is necessary.

Now we turn to non-equilibrium processes. As an extreme case, we turn on the coupling instantaneously.  The second protocol remains zero during the first period ($t_1>t>t_0$) but it is identical to protocol 1 for the remaining periods.  Since  $V_\sI=0$, no work is done by the sudden change of $\lambda$.  Between $t_1$ and $t_2$, the system relaxes to an equilibrium, during which dissipation takes place. The entropy production can be measured at $t_2$ and $t_4$. At $t_4$, $\Delta \mathcal{F}=0$ and $\Delta \mathcal{S}=0$ and thus the net dissipation is given by 
\begin{equation}\label{eq:sigma_W4}
   \Sigma_\sW(t_4)  = \beta \left [W(t_4) - \Delta U(t_4)\right], 
\end{equation}
\begin{equation}\label{eq:sigma_Q4}
\Sigma_\sQ(t_4) = -\beta Q(t_4).
\end{equation}
which include the initial dissipation $\Sigma^0$.

We make another measurement at $t_2$ where the relaxation has completed.  Since no work is done before this point,  the amount of dissipation is just the free energy difference:
\begin{equation}\label{eq:sigma_W2}
   \Sigma_\sW(t_2) =  W(t_2) - \Delta \mathcal{F} = \beta \left [- W^\text{qs}(t_2) + W^\text{qs}(t_4) - \Delta U^\text{qs}(t_4)\right].
\end{equation}
On the other hand, heat flows during the relaxation. The dissipation obtained from heat is
\begin{equation}\label{eq:sigma_Q2}
   \Sigma_\sQ(t_2) =  \Delta \mathcal{S} - \beta Q(t_2) = \beta \left[ Q^\text{qs}(t_2) - Q^\text{qs}(t_4) - Q(t_2) \right].
\end{equation}
We have used only the standard theory of thermodynamics based on the transition between equilibrium states and thus these expressions (\ref{eq:sigma_W4}) -- (\ref{eq:sigma_Q2}) should be correct and they all coincide. These measured quantities serve as a benchmark test of the computational experiment.

Now, we try to extend the expression of dissipation to a ``non-equilibrium'' expression (time-dependent expression) for the period where dissipation is actually taking  place ($t_2>t>t_1$).  Since no work is done during this period, time-dependent dissipation based on work is not possible.
Therefore, we determine the dissipation from heat. By definition, the dissipative heat is  $Q^\text{rev}(t)-Q(t)$ where the reversible heat is $Q_\text{rev}(t) = Q^\text{qs}(t) - Q^\text{qs}(t_4)$ in the current experimental setting. It is reasonable to define time-dependent entropy production as
\begin{equation}\label{eq:sigma_Q}
   \Sigma_\sQ(t) = \beta \left [Q^\text{qs}(t) - Q^\text{qs}(t_4) - Q(t) )   \right], \qquad t > t_1.
\end{equation}
While this expression takes the correct values at the two equilibrium measurement points, its validity is not certain since the coupling energy also deviates from its equilibrium value during this period.  

ELV has proposed a promising expression\cite{Esposito2010,Funo2019}: 
\begin{equation}\label{eq:sigma_S}
\Sigma_S(t) = \Delta S_\sS(t) - \beta Q(t)
\end{equation}
which is strictly positive.
It replaces the reversible entropy change $\beta \left[Q^\text{qs}(t)-Q^\text{qs}(t_4)\right]$ in Eq. (\ref{eq:sigma_Q}) with the change in vN-entropy $\Delta S_\sS(t) = - \tr_\sS \rho_\sS(t)  a\ln \rho_\sS(t) + \tr_\sS \rho_\sS(t_0) \ln \rho_\sS(t_0)$. 
One important feature of this expression is that the dissipation due to decoherence is explicitly included.   It has been also claimed that this expression is valid even when the coupling is strong. However, it is not obvious.   

In next section, we construct a simple model  and measure the \emph{conventional} entropy production from expressions (\ref{eq:sigma_W4})--(\ref{eq:sigma_Q2}), which serves as a benchmark for the proposed theory. The two \emph{non-equilibrium} expressions of entropy production (\ref{eq:sigma_Q}) and (\ref{eq:sigma_S}) are evaluated and their validity will be checked.

\section{Model}

\begin{figure}
   \begin{subfigure}{0.45\textwidth}
      \centering
      \includegraphics[height=4cm]{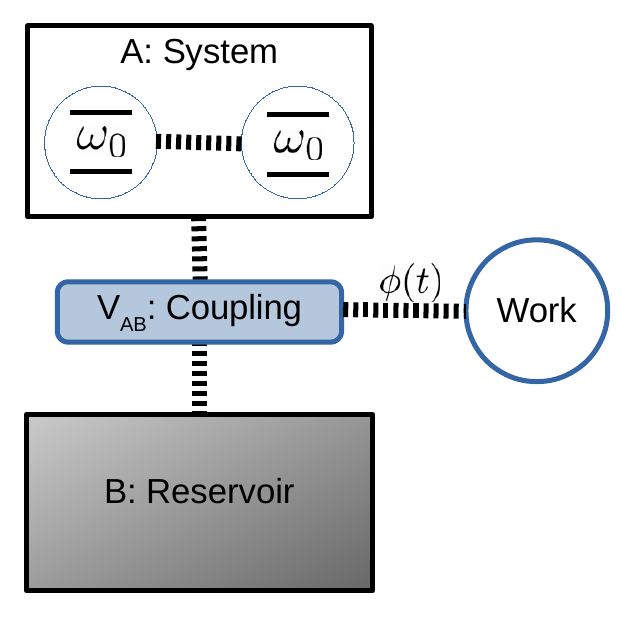}
      \caption{A pair of qubits in a system S is connected and disconnected to an environment through the time-dependent protocol $\lambda(t)$.  During transition periods, work is done through the time-varying coupling.}
      \label{fig:model}
   \end{subfigure}
   \hspace{0.5cm}
   \begin{subfigure}{0.45\textwidth}
      \centering
      \includegraphics[height=4cm]{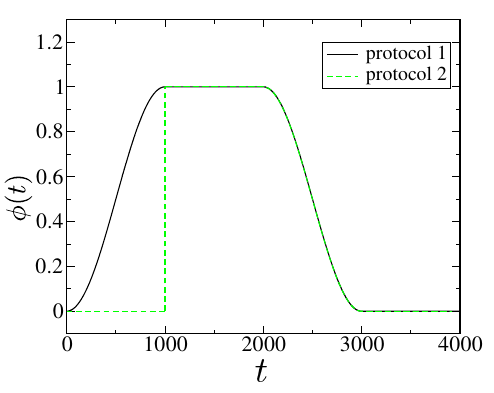}
      \caption{Two protocols. The coupling is slowly turned on in protocol 1 and instantaneously in protocol 2.  In both protocols, the coupling is slowly turned off.} 
      \label{fig:protocols}
   \end{subfigure}
   \caption{Model}   
\end{figure} 

We consider a simple spin-Boson model in which a pair of coupled qubits interact with an environment consisting of an ideal Bose gas. (See Fig. \ref{fig:model}.) Hamiltonian of the system is given by
\begin{equation}
   H_\sS = \frac{\omega_0}{2} \sigma_{z}\otimes I+\frac{\omega_0}{2} I\otimes \sigma_{z} + \Lambda (\sigma_{+}\otimes \sigma_{-}+
   \sigma_{-}\otimes \sigma_{+})
\end{equation}
where $\omega_0$ and $\Lambda$ is the excitation energy of individual qubits and a coupling strength between them, respectively.   We use $\omega_0=1$ and $\Lambda=0.5$ for all case studies. (All energies are scaled with $\omega_0$ through out the paper.)

The environment is an ideal Bose gas with Hamiltonian
\begin{equation}
   H_\sB = \sum_{j\ge 1} \omega_j a^\dagger_j a_j
\end{equation}
and it interacts with the system through a bilinear form of coupling Hamiltonian
\begin{equation}
   H_\sI(t) = \lambda(t)\, X_\sS \otimes Y_\sB
\end{equation}
where the strength of coupling is controlled by a protocol $0 \le \lambda(t) \le 1$.
The environment side of the bilinear coupling  is a displacement of the environment
\begin{equation}
   Y_\sB = \sum_j \nu_j \left(a^\dagger_j + a_j \right).
\end{equation}
with coupling strength $\nu_j$ between the system and $j$-th mode.
The spectral density of the environment is assumed to be of the Drude-Lorentz type
\begin{equation}
   J(\omega) = \frac{2\kappa}{\pi} \frac{\omega\gamma}{\omega^2+\gamma^2}
\end{equation}
where $\kappa$ and $\gamma$ are the coupling strength and relaxation rate, which are fixed at $\kappa=1$ and $\gamma=0.1$ in the present investigation.
For the system side of the coupling operator,  we will explore various different $x_\sS$.   

We assume the the whole system is completely isolated and its unitary evolution is determined by the Liouville -- von Neumann equation, $\dv{t}\rho_\sSB = -i \comm{H_\sSB}{\rho_\sSB}$.
Taking partial trace, the reduced density follows a non-unitary evolution 
\begin{equation}\label{eq:eom}
   \dv{t} \rho_\sS = -i \comm{H_\sS}{\rho_\sS} - i \lambda(t) \comm{X_\sS}{\eta_\sS}
\end{equation}
where $\eta_\sS = \tr_\sB \left(Y_\sB \rho_\sSB \right)$ is an operator in the system Hilbert space. 
If the initial state is a product state $\rho_\sSB(t_0) = \rho_\sS(t_0) \otimes \rho_\sB^\sG$, then Eq (\ref{eq:eom}) can be numerically solved using the method of hierarchical equations of motion (HEOM) \cite{Tanimura1990,Kato2015} and the results are in principle exact.  A breif summary of HEOM is given in Appendix.
The operator $\eta_\sS$ contains sufficient information about the state of the environment and allows us to evaluate various thermodynamics quantities.  

Heat (\ref{eq:def_Q}), work (\ref{eq:def_W}), and coupling energy (\ref{eq:def_VI}) are computed as
\begin{equation}
   Q = - \int_{t_0}^{t} \left ( \tr_\sS\left[H_\sS \dot{\rho}_\sS(\tau)\right]+\lambda(\tau) \tr_\sS\left[X_\sS \dot{\eta}(\tau)\right] \right ) \dd{\tau},
\end{equation}
\begin{equation}
W =\int_{t_0}^{t} \dot{\lambda}(\tau) \tr_\sS \left\{X_\sS \eta(\tau)\right\} \dd{\tau},
\end{equation}
and
\begin{equation}
   V_\sI = \lambda(t) \tr_\sS \left \{X_\sS \eta_\sS(t)\right\},
\end{equation}
which satisfy the conservation of energy $W + Q = \Delta U + \Delta V_\sI$.
In addition, we monitor the presence of system-environment correlation by computing correlation function
\begin{equation}
   C_\sSB(t) \equiv \expval{X_\sS \otimes Y_\sB} - \expval{X_\sS} \expval{Y_\sB} = \tr_\sS\left(X_\sS \eta_\sS\right) - \tr_\sS\left( X_\sS \rho_\sS\right)\, \cdot\tr_\sS\left(\eta_\sS\right).
\end{equation}

\begin{figure}
   \centering
   \includegraphics[width=0.9\textwidth]{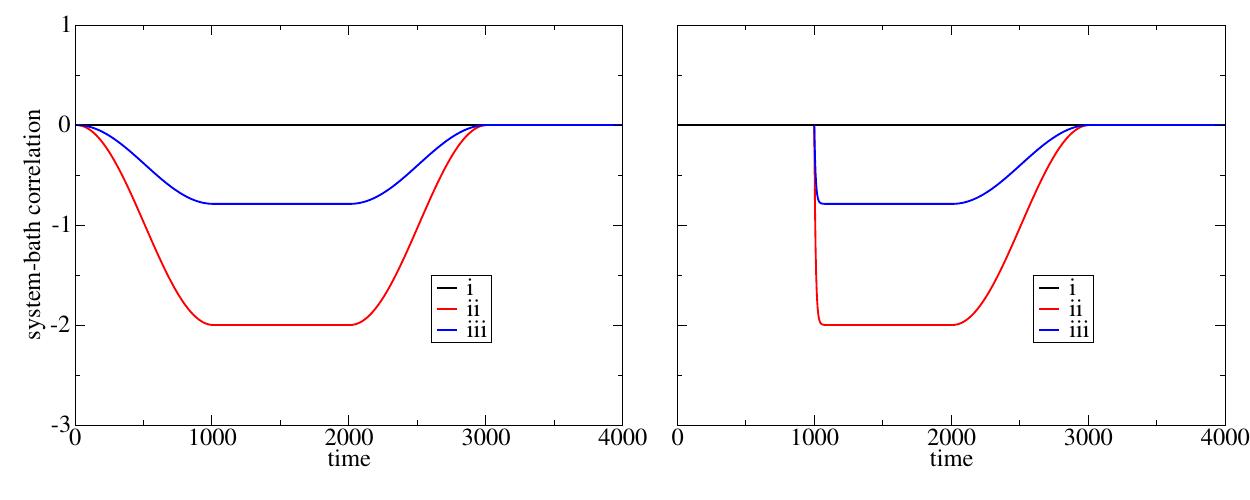}
   \caption{Correlation between system and environment for the three initial states given in Eq (\ref{eq:rho_ini}) during protocol 1 (left) and protocol 2 (right).  Initial state (\romannumeral 1) shows no correlation at all. (\romannumeral 2) and (\romannumeral 3) forms quantum and classical correlation, respectively.  The formation of correlation is rather quick as shown in the right panel.}\label{fig:correlation}
\end{figure}

The quasi static process is emulated by protocol 1:
\begin{equation}
   \lambda_1(t) = 
   \begin{cases}
      \sin^2\left[\frac{\pi t}{2 \tau} \right] & 0 < t < \tau \\
      1 & \tau < t < 2 \tau \\
      \sin^2\left[\frac{\pi(3\tau - t)}{2\tau}\right] & 2\tau < t < 3\tau\\
      0 & 3 \tau < t < 4 \tau
   \end{cases}
\end{equation}
where $\tau=1000$. (See Fig. \ref{fig:protocols}.)  For the relaxation experiment (protocol 2), $\lambda_2(t)= 0$ for $t_1> t > t_0$ and otherwise $\lambda_2(t)=\lambda_1(t)$.  This allows us to compare equilibrium and non-equilibrium states easily on the plots.

\section{Case studies}

We choose three particular initial states: 
\begin{equation}\label{eq:rho_ini}
\rho_\sS(t_0) = 
\begin{cases}
\text{(\romannumeral 1)}& \dyad{00}\\
\text{(\romannumeral 2)}& \frac{1}{2} \left (\dyad{00}+\dyad{00}{11}+\dyad{11}{00}+\dyad{11}{11}\right) \\
\text{(\romannumeral 3)}& e^{-\beta H_\sS}/\tr\left(e^{-\beta H_\sS}\right)
\end{cases}
\end{equation}
which allow us to control S-B correlation. 

First we consider a special choice of coupling $X_\sS=H_\sS$. Since $H_\sS$ commutes with the total Hamiltonian, the energy of system conserves, i.e., $\Delta U=0$ which highlights the effect of coupling energy since heat released by the environment never enters the system.  The diagonal elements of the reduced density $\rho_\sS$ in the energy basis are invariant but the off-diagonal elements vanish  due to decoherence as soon as the interaction with environment is turned on. We also consider a more general form of coupling, $X_\sS = \sigma_x \otimes I + I \otimes \sigma_x$.

\begin{figure}
   \begin{subfigure}[t]{0.45\textwidth}
   \centering
   \includegraphics[width=2.0in]{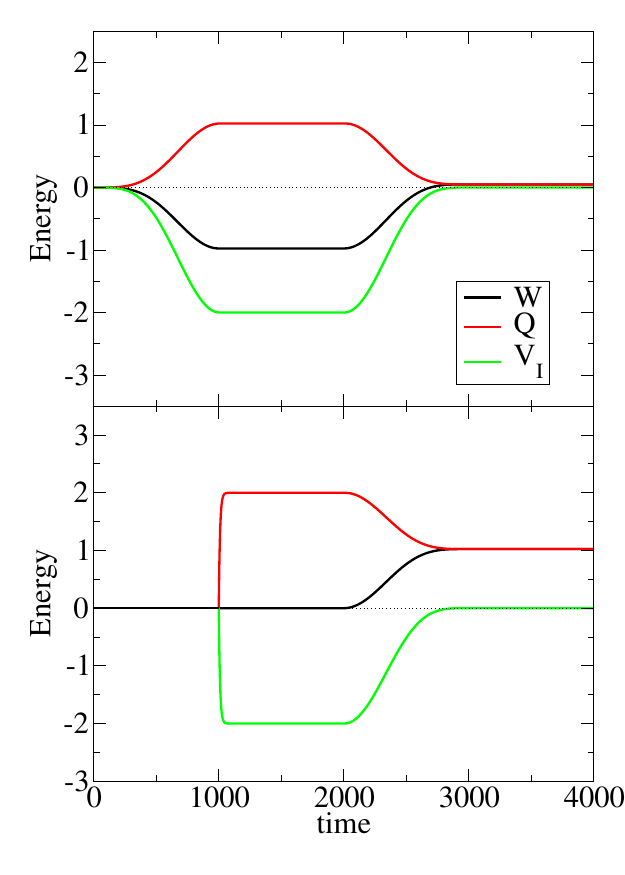}
   \caption{Energy transaction during protocol 1 (top) and 2 (bottom).}\label{fig:case_a1}
   \end{subfigure}
   \hspace{0.1in}
   \begin{subfigure}[t]{0.45\textwidth}
   \centering
   \includegraphics[width=2.0in]{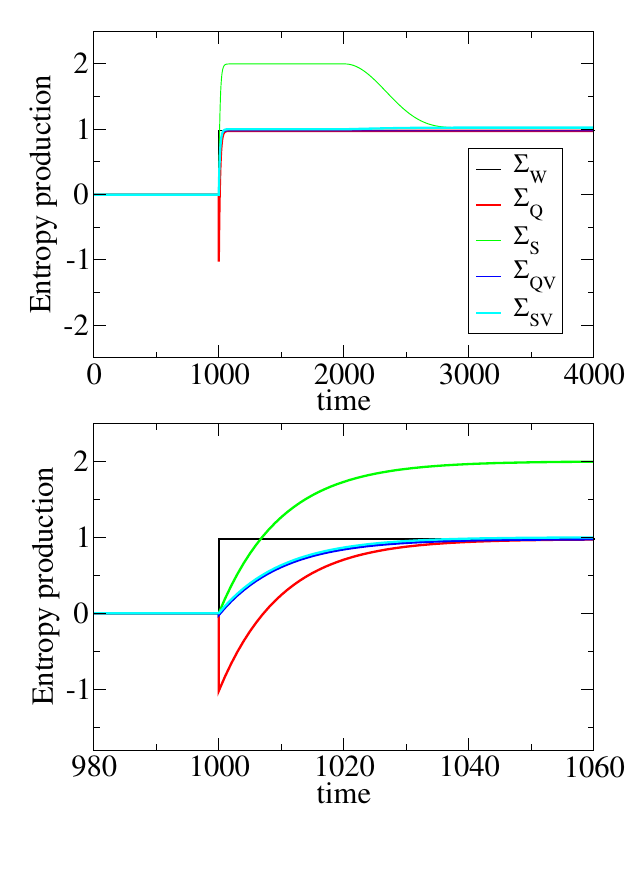}
   \caption{Various definitions of entropy production evaluated from the computer experiment.  Whole process (top) and during the relaxation process takes place (bottom).}\label{fig:case_a2}
   \end{subfigure}
\caption{Time evolution various thermodynamics quantities for case I.}
\end{figure}

\subsection{Case I}

Initial state (\romannumeral 1) is an energy eigenstate, which is protected from the environment by $X_\sS = H_\sS$.  Since the reduced density remains in the pure state, no correlation of any kind is formed through out the time-evolution.
Figure \ref{fig:correlation} confirms that there is no S-B correlation and thus the whole system remains in a product state.  The entropy production takes place only in the environment.  Figure \ref{fig:case_a1} plots the energy changes. For protocol 1, both work and heat  vanish at the end, certifying the process is quasi static.  For protocol 2, 
both work and heat reach the same value at $t_4$.  Using Eqs. (\ref{eq:sigma0_Q4}) and (\ref{eq:sigma0_W4}),  we find $\Sigma_\sQ(t_4) = \Sigma_\sW(t_4) = 1.0$, which is the dissipation incurred during initial relaxation.  At the other check point $t_2$, we obtain entropy production $\Sigma_\sQ(t_2) = \Sigma_\sW(t_2) = 0.98$ from Eqs. (\ref{eq:sigma_W2}) and (\ref{eq:sigma_Q2}).  The slight difference between the two check points is due to the small dissipation caused by the finite time protocols.  Based on these observations, we conclude that the simulation produces a standard thermodynamic behavior of the quasi static and relaxation processes.

Next, we plot non-equilibrium expressions (\ref{eq:sigma_Q}) and (\ref{eq:sigma_S}) in Fig. \ref{fig:case_a2}. After the rapid relaxation period, $\Sigma_\sQ$ reaches the correct dissipation and passes through the two check points.  On the other hand, $\Sigma_\sS$ shows disagreement.  While the correct value is reached at the end, it grossly overestimates the dissipation at the check pint $t_2$. As mathematically proved, $\Sigma_S$ is always positive, consistent with the second law but its derivative is clearly negative between $t_2$ and $t_3$.    Hence, $\Sigma_\sS$ cannot be considered as a valid definition of the entropy production.  Zooming into the period where the dissipation actually takes place (the bottom panel of Fig.\ref{fig:case_a2}), $\Sigma_\sQ$ initially goes down below 0, which is incompatible with the second law. We found that  neither $\Sigma_\sQ$ nor $\Sigma_\sS$ is consistent with the second law.

However, we have noticed that if the entropy is redefined as $S(t) \rightarrow S(t) + \frac{1}{2}V_\sI(t)$, both $\Sigma_\sQ$ and $\Sigma_\sS$ coincide and become consistent with the second law.  Based on these observations, we introduce two empirical expressions of entropy production:
\begin{equation}
   \Sigma_\sQV(t) = \Sigma_\sQ(t) + \frac{1}{2} \left[V_\sI^\text{qs}(t) - V_\sI(t)\right]
\end{equation}
and 
\begin{equation}
   \Sigma_\sSV(t) = \Sigma_\sS(t) + \frac{1}{2} \Delta V_\sI(t)
\end{equation}
where $V_\sI^\text{qs}(t)$ and $V_\sI(t)$ are the coupling energy along protocol 1 and 2, respectively.
Figure \ref{fig:case_a2} show that $\Sigma_\sQV$ and $\Sigma_\sSV$ both monotonically increase to the correct value as expected from the second law.

\begin{figure}
   \begin{subfigure}[t]{0.45\textwidth}
      \centering
      \includegraphics[width=2.0in]{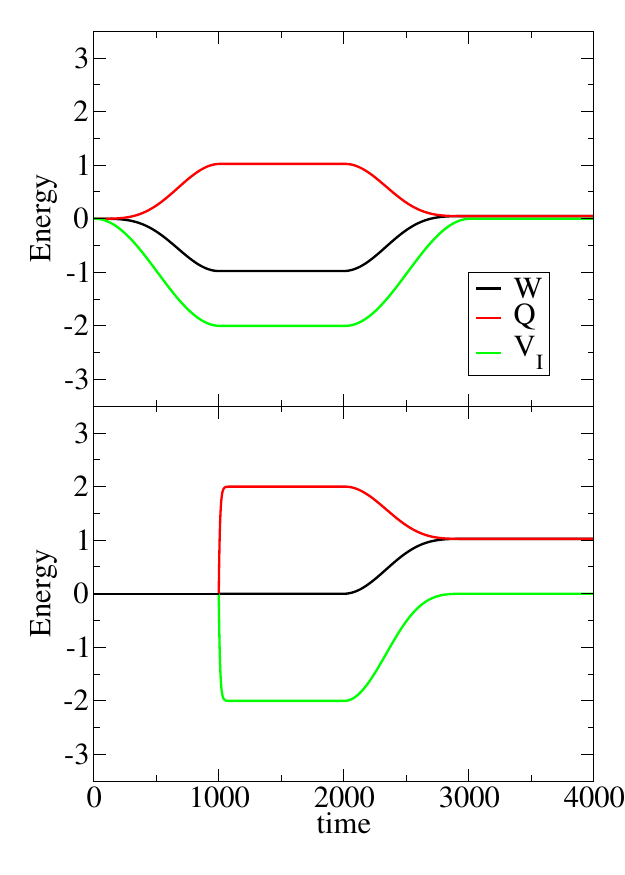}
      \caption{Energy transaction during protocol 1 (top) and 2 (bottom).}\label{fig:case_b1}
   \end{subfigure}
   \hspace{0.1in}
   \begin{subfigure}[t]{0.45\textwidth}
      \centering
      \includegraphics[width=2.0in]{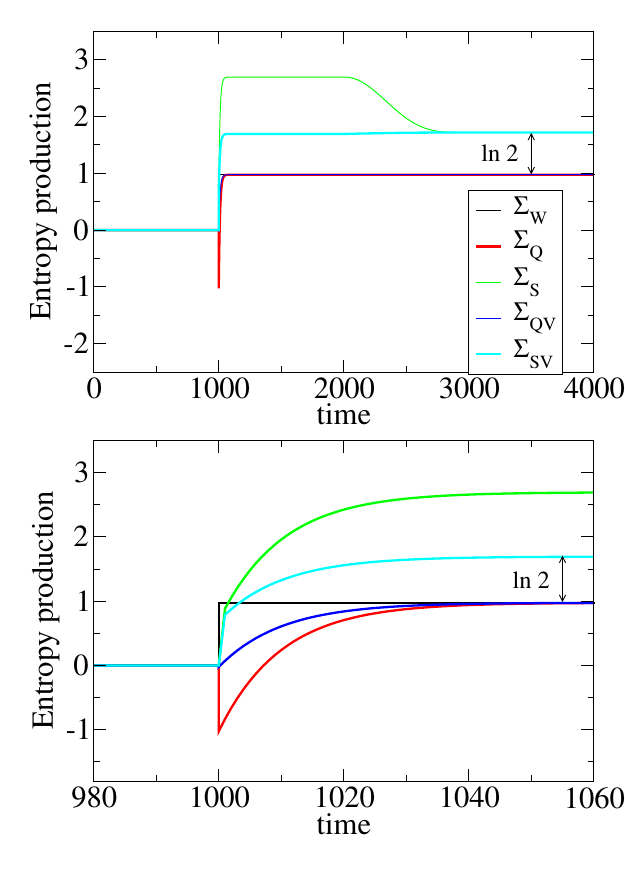}
      \caption{Various definitions of entropy production evaluated from the computer experiment.  Whole process (top) and during the relaxation process takes place (bottom).}\label{fig:case_b2}
   \end{subfigure}
   \caption{Time evolution of various thermodynamics quantities for case II.}
\end{figure}

\subsection{Case II}
The coupling operator $X_\sS$ is still $H_\sS$.
Initial state (\romannumeral 2) is a superposition of two energy eigenstates, which is subject to decoherence due to quantum entanglement with the environment. The off-diagonal elements in the initial pure state vanish and $S_\sS$ increases. 

Figure \ref{fig:correlation} indeed shows a strong S-B correlation, which is presumably due to entanglement. In terms of energy transaction, this case is almost identical to case I, and thermodynamic dissipations measured at two check points are consistent (Fig. \ref{fig:case_b1}).
The entropy production evaluated from heat, $\Sigma_\sQ$ failed in the same way as in the previous case.  Similarly, $\Sigma_\sS$ failed as well. Applying the same empirical correction introduced in case I, $\Sigma_\sQV$ monotonically increases to the value obtained from $\Sigma_\sW(t_4)$ and $\Sigma_\sQ(t_4)$.  Hence, it is consistent with thermodynamics.  On the other hand, $\Sigma_\sSV$ reaches a higher value due to the decoherence-induced entropy production.
When the decoherence is perfect, $\ln 2$ of irreversible entropy is expected.  Figure \ref{fig:case_b2} shows that $\Sigma_\sSV-\Sigma_\sQV = \ln 2$ in good agreement.

\subsection{Case III}

We still keep the same $X_\sS$. The system starts with a Gibbs state (\romannumeral 3) with the same temperature as the environment and thus the whole system is in a thermal equilibrium at the beginning.  It is a classical mixture of the energy eigenstates protected from the environment. Thus only the classical correlation between the system and environment can be formed and the $S_\sS$ remains constant. 
Figure \ref{fig:correlation} shows a weaker correlation than the entanglement in case II.  Since the initial state is diagonal, there is no decoherence and thus $\Sigma_\sQV$ and $\Sigma_\sSV$ are nearly identical. Both $\Sigma_\sQ$ and $\Sigma_\sS$ fails as the previous cases and the empirical expressions work for this case as well as the previous cases.

\begin{figure}
   \begin{subfigure}[t]{0.45\textwidth}
      \centering
      \includegraphics[width=2.0in]{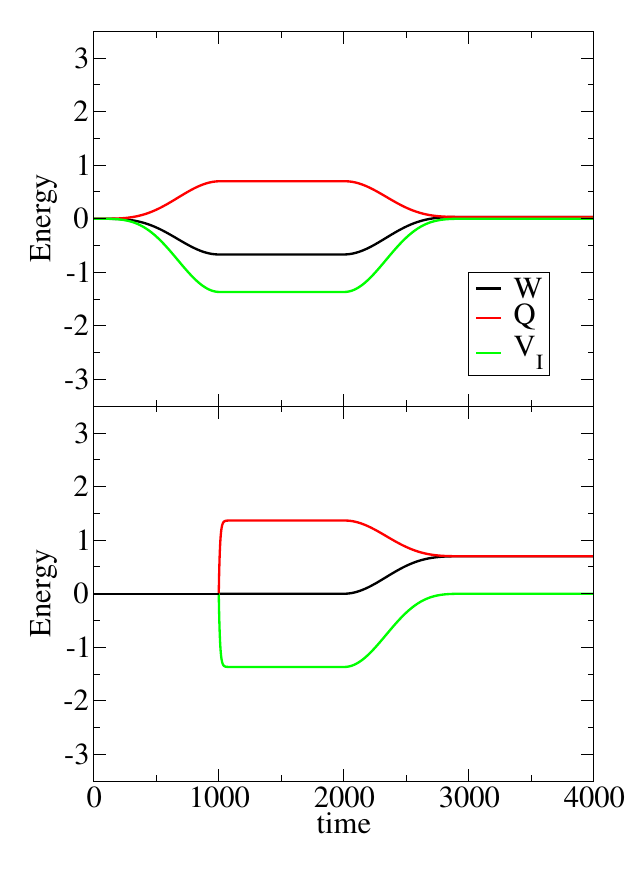}
      \caption{Energy transaction during protocol 1 (top) and 2 (bottom).}\label{fig:case_c1}
   \end{subfigure}
   \hspace{0.1in}
   \begin{subfigure}[t]{0.45\textwidth}
      \centering
      \includegraphics[width=2.0in]{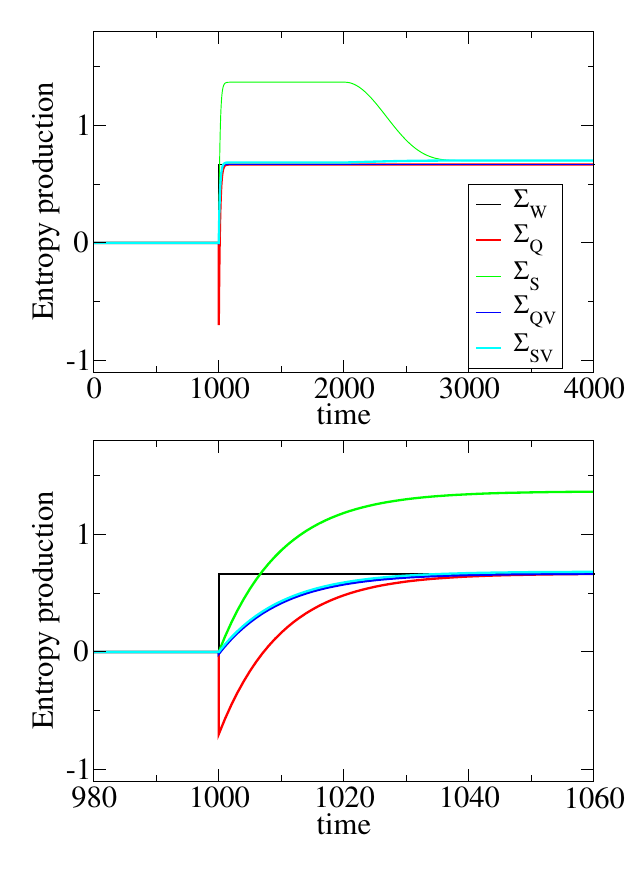}
      \caption{Various definitions of entropy production evaluated from the computer experiment.  Whole process (top) and during the relaxation process takes place (bottom).}\label{fig:case_c2}
   \end{subfigure}
   \caption{Time evolution various thermodynamics quantities for case III.}
\end{figure}

\subsection{Case IV}

\begin{figure}
   \begin{subfigure}[t]{0.45\textwidth}
      \centering
      \includegraphics[width=2.0in]{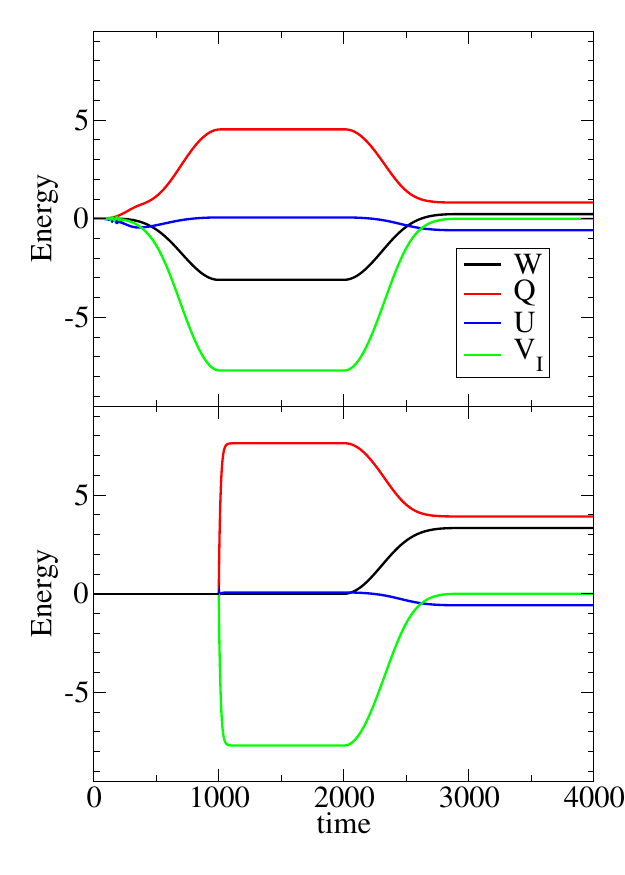}
      \caption{Energy transaction during protocol 1 (top) and 2 (bottom).}\label{fig:case_d1}
   \end{subfigure}
   \hspace{0.1in}
   \begin{subfigure}[t]{0.45\textwidth}
      \centering
      \includegraphics[width=2.0in]{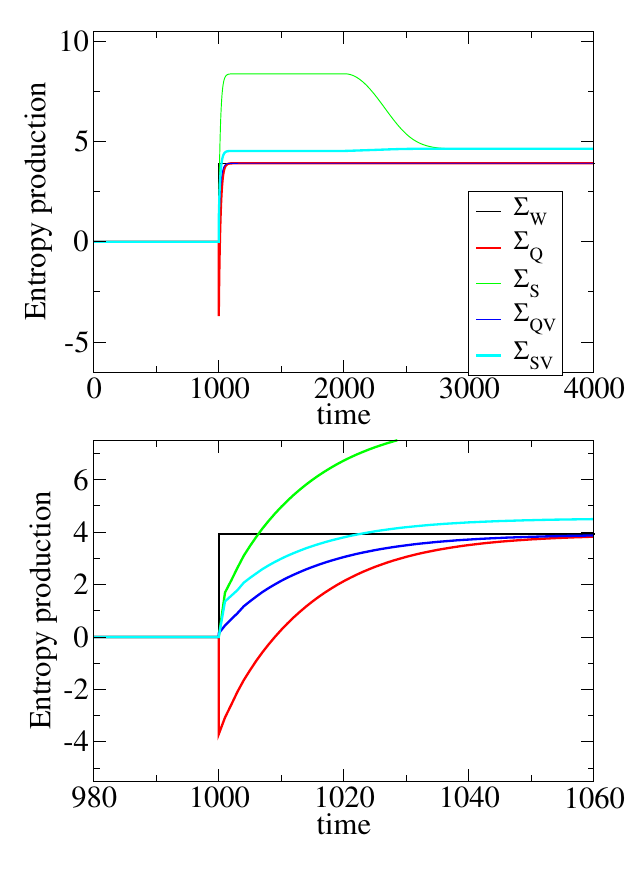}
      \caption{Various definitions of entropy production evaluated from the computer experiment.  Whole process (top) and during the relaxation process takes place (bottom).}\label{fig:case_d2}
   \end{subfigure}
   \caption{Time evolution various thermodynamics quantities for case IV.}
\end{figure}

We consider a more realistic coupling  $X_\sS = \sigma_x \otimes I + I \otimes \sigma_x$. Due to the symmetry between the two qubits, the singlet state (decoherence free state) is protected from the environment.  The remaining triplet states are thermalized.   Unlike the previous special cases, the system energy relaxes and thus some portion of heat flows into the system. Starting with the non-equilibrium initial state (\romannumeral 2), we expect both classical and quantum correlations are formed.  The energy transaction plotted in Fig. \ref{fig:case_d1} are a little more complicated than the previous cases. Figure \ref{fig:case_d2} shows that $\Sigma_\sQ$ and $\Sigma_\sS$ fail in the same way as before.  Remarkably, both $\Sigma_\sQV$ and  $\Sigma_\sSV$ still behave reasonably and  seem consistent with the second law despite of energy exchange between the system and the environment.  The dissipation due to the decoherence is clearly visible in $\Sigma_\sSV - \Sigma_\sQV$.

\section{Discussion}

We found that the entropy production $\Sigma_S$  proposed in Ref. \cite{Esposito2010} is not consistent with the second law as it is. The conventional interpretation of dissipation defined by Eq. (\ref{eq:sigma_Q}) also failed.  However, the empirical redefinition of entropy made them more consistent.  In order to justify the correction, we briefly review the more recent theory based on the potential of mean force.\cite{Gelin2009,Campisi2009,Campisi2010,Hilt2011,Seifert2016,Jarzynski2017,Miller2017,Strasberg2019}
Here we use the quantum version based on Ref. \cite{Strasberg2019}.

First the theory assumes that the whole system is in a canonical equilibrium and defines a temperature $\beta$ for a given energy by
\begin{equation}
   E_\text{total} = \tr_\sSB \left \{(H_\sS+H_\sB+V_\sSB)e^{-\beta(H_\sS+H_\sB+V_\sSB)}\right\}/Z_\sSB
\end{equation}
where $Z_\sSB = \tr_\sSB \left \{e^{-\beta(H_\sS+H_\sB+V_\sSB)} \right\}$.  Assuming that the reference temperature does not change over time, an effective Hamiltonian\cite{Strasberg2019} for the system is defined by
\begin{equation}
   \tilde{H}_\sS = -\frac{1}{\beta} \ln \left[ \tr_\sB\left \{e^{-\beta(H_\sS+H_\sB+V_\sSB)} \right\}/Z_\sB\right]
\end{equation}
with $Z_\sB=\tr_\sB e^{-\beta H_\sB}$.
Further, the effective thermodynamics quantities are redefined for an arbitrary state of system $\rho_\sS(t)$ by
\begin{equation}
   \tilde{U} = U +  \tilde{U}' + \tilde{U}''
\end{equation}
\begin{equation}
   \tilde{S} = S_\sS + \beta \tilde{U}''
\end{equation}
\begin{equation}
   \tilde{F} = F +  \tilde{U}'
\end{equation}
\begin{equation}
   \tilde{Q} = Q  - \Delta V_\sSB + \tilde{U}' + \Delta \tilde{U}''
\end{equation}
where the additive correction terms are defined by
\begin{subequations}
   \begin{equation}
      \tilde{U}'= \tr_\sS \{\rho_\sS(t)(\tilde{H}_\sS-H_\sS)\}
   \end{equation}
   \begin{equation}
      \tilde{U}'' = \tr_\sS \{\rho_\sS(t) \tilde{\beta} \partial_{\beta} \tilde{H}_\sS\}.
   \end{equation}
\end{subequations}
These effective quantities satisfy usual thermodynamic relations.
Using these definitions along with the first law $W+\tilde{Q}=\Delta \tilde{U}$ we find an expression of entropy production 
\begin{equation}\label{eq:secondlaw_mf}
   \tilde{\Sigma} = \Sigma_\sS + \beta \tilde{U}'
\end{equation}

The present empirical correction is equivalent to $\tilde{U}' = \tilde{U}''=\frac{1}{2}V_\sI$.  Then, the effective thermodynamic quantities become
\begin{subequations}
   \begin{equation}
      \tilde{U} = U +  V_\sI
   \end{equation}
   \begin{equation}
      \tilde{S} = S_\sS + \frac{1}{2}V_\sI
   \end{equation}
   \begin{equation}
      \tilde{F} = F + \frac{1}{2}V_\sI
   \end{equation}
   \begin{equation}
      \tilde{Q} = Q
   \end{equation}
   \begin{equation}
   \tilde{\Sigma} = \Sigma_\sS + \frac{1}{2} \Delta V_\sI.
\end{equation}
\end{subequations}

Admittedly  we are not able to derive the correction terms from the first principle. We do not claim that they are exact.  However, the additive corrections obtained from the potential of mean force is quite consistent with our finding.   The correction terms in the mean force theory do not depend on the state of environment whereas the coupling energy does.  So, the corresponding between the mean force theory and our results are still not clear.  Further theoretical investigation is needed.

\section{Conclusion}

We have developed an exact numerical experiment of qubits strongly coupled to a thermal environment.  Information about equilibrium states is obtained from a slow protocol mimicking a quasi static process.  The dissipation during relaxation processes is measured  based on the standard thermodynamic principle.  Based on the entropy production obtained from the experiment and the standard relation between entropy and heat, we defined a time-dependent expression of entropy production consistent with the observed value.    Then, we tested it along with a theoretical expression. Neither was found to be consistent with the second law.  However, a simple additive correction to entropy makes them consistent with the second law.  For the current model, the correction is a half of the coupling energy.  Accordingly, other quantities need to be adjusted to satisfy the 1st and 2nd laws.  The results are qualitatively consistent with the theory based on the potential of mean force. 

\section*{Acknowledgments}

We gratefully acknowledge the stimulating discussion with Janet Anders on the method of effective Hamiltonian method and quantum thermodynamics in general.  My deepest and sincere appreciation goes to late Christian Van den Broeck  who taught me non-equilibrium statistical mechanics for thirty years.
 
\appendix
\section{Hierarchical Equations of Motion}
Kato and Tanimura \cite{Kato2015} calculated heat conduction with essentially the same model using the same method of hierarchical equation of motion (HEOM) as the present work. Here only the equations we used are shown without derivation.

The pair correlation function are approximated as
$\expval{Y_\sB(\tau)Y_\sB(0)}=\lambda \left(c_1 e^{-\gamma_1 \tau}+c_2 e^{-\gamma_2 \tau} + 2 c_0 \delta(\tau) \right)$. The coefficients $c_j$ and the decay rates $\gamma_j$ are fitted to the Drude-Lorentz spectral density. If the whole system is initially in a product  state $\rho_\sS(t_0) \otimes \rho_\sB^\sG$, the non-unitary equation of motion (\ref{eq:eom}) can be mapped to a hierarchy of Markovian equations: 
\begin{eqnarray}\label{eq:HEOM-schroedinger}
   \dv{t}\zeta_{n_1,n_2}(t) &=& -i [H_\sS,\zeta_{n1,n2}]_{-} \nonumber \\
   &&  - (\gamma_1 n_1 + \gamma_2 n_2) \zeta_{n1,n2}(t) - \lambda c_0 \lambda^2(t)\mathcal{S}^{-} \mathcal{S}^{-}\, \zeta_{n1,n2}(t) \nonumber \\
   && -i n_1 \lambda(t) \mathcal{G}_1\, \zeta_{n_1-1,n_2}(t) -i n_2 \lambda(t) \mathcal{G}_2\, \zeta_{n_1,n_2-1}(t) \nonumber \\
   && -i \lambda \lambda(t) \mathcal{S}^{-} \left\{\zeta_{n_1+1,n_2}(t) + \zeta_{n1,n2+1}(t) \right\}
\end{eqnarray} 
where $\zeta_{n_1,n_2}$ are auxiliary operators.   Super operators $\mathcal{S}^\pm = \comm{X_\sS}{\cdot}_\pm$ are commutator and anti-commutator. Another super operator is defined by 
\begin{equation}
   \mathcal{G}_j = \Re\{c_j\} \mathcal{S}^{-} + i \Im\{c_j\}\mathcal{S}^{+} .
\end{equation}

The system density $\rho_\sS$ and the moment operator $\eta_\sS$ are obtained from the auxiliary operators as
$\rho_\sS(t) = \zeta_{0,0}(t)$ and 
\begin{equation}
   \eta_\sS(t) = \lambda(t) \left[ \zeta_{1,0}(t)+\zeta_{0,1}(t) + i c_0 \mathcal{S}^{-}\zeta_{0,0}(t) \right ].
\end{equation}
Although we need only top three auxiliary operators, they are tightly coupled to higher depths especially when non-Markovian effects are strong.  For a system strongly coupled with environments, we must include many auxiliary operators of higher depth.
The hierarchy is terminated at a certain depth $d$ depending on the coupling strength.  In the present case studies, $d=50$ is used. 
 
\section*{References}

\bibliographystyle{elsarticle-num}      
\bibliography{refs}

\end{document}